\documentstyle[12pt]{article}
\begin{document}

\def\pp{{\, \mid \hskip -1.5mm =}}
\def\cL{{\cal L}}
\def\be{\begin{equation}}
\def\ee{\end{equation}}
\def\bea{\begin{eqnarray}}
\def\eea{\end{eqnarray}}
\def\beaa{\begin{eqnarray*}}
\def\eeaa{\end{eqnarray*}}
\def\tr{{\rm tr}\, }
\def\nn{\nonumber \\}
\def\e{{\rm e}}
\def\D{{D \hskip -3mm /\,}}

\  \hfill 
\begin{minipage}{3.5cm}
OCHA-PP-189 \\
April 2002 \\
\end{minipage}

\vfill

\begin{center}
{\large\bf Stabilization and radion  in de Sitter brane-world}

\vfill

{\sc Shin'ichi NOJIRI}\footnote{nojiri@cc.nda.ac.jp}, 
{\sc Sergei D. ODINTSOV}$^{\spadesuit}$\footnote{
odintsov@mail.tomsknet.ru}, \\
and {\sc Akio SUGAMOTO}$^{\heartsuit}$\footnote{
sugamoto@phys.ocha.ac.jp}
\\

\vfill

{\sl Department of Applied Physics \\
National Defence Academy, 
Hashirimizu Yokosuka 239, JAPAN}

\vfill

{\sl $\spadesuit$
Lab. for Fundamental Study, 
Tomsk State Pedagogical University, 634041 Tomsk, RUSSIA}

\vfill

{\sl $\heartsuit$ Department of Physics, 
Ochanomizu University \\
Otsuka, Bunkyou-ku Tokyo 112, JAPAN}

\vfill

{\bf ABSTRACT}

\end{center}
We consider the stabilization of de Sitter brane-world. 
The scalar field bulk-brane theory produces the non-trivial minimum 
of modulus potential where temporal radion is realized. The  
hierarchy problem (between Planck and electroweak scales) may 
be solved.
However, the interpretation of radion is not so clear as in AdS brane-world.
In particulary, the introduction of two times physics or pair-creation of
bulk spaces or identification of one of spatial coordinates with imaginary
time (non-zero temperature)  may be required.

\vfill

\noindent
PACS: 98.80.Hw,04.50.+h,11.10.Kk,11.10.Wx

\noindent
Keywords: radion, dS/CFT, brane world

\newpage

The investigation of five-dimensional brane-worlds naturally posed the
problem of stabilization of extra dimension. In case of AdS bulk the
corresponding mechanism of stabilization has been proposed by
Goldberger-Wise\cite{wise}
via the realization of the minimum of the corresponding modulus effective
potential. However, the explicit stabilization of fifth dimension (without 
fine-tuning) may be presumably realized only by thermal quantum effects 
\cite{milton}. The  physics of radion (which vacuum expectation value
coincides with the fifth dimension radius) may play an important role in
electroweak theory where , for example, S-parameter turns out to be modified.

The interest to AdS brane-worlds is motivated mainly by AdS/CFT
correspondence. However, it has been recently proposed dS/CFT 
correspondence (see \cite{strominger}for an introduction).
As a result it is naturally to search for realistic dS brane-world scenarios.
As one step in this direction we study how radion physics (or stabilization of
extra dimension) may be realized in dS brane-world.

Let us start from the 5-dimensional de Sitter space 
as a bulk spacetime:
\be
\label{i}
ds^2 = G_{\mu\nu}dx^\mu dx^\nu 
= G_{tt}dt^2 + \sum_{i,j=1}^4g_{ij}dx^i dx^j
= - dt^2 + \e^{-2k|t|}\sum_{i=1}^4\left(dx^i\right)^2\ .
\ee
We also consider the scalar field, whose action is given by
\be
\label{ii}
S_\phi=\int\sqrt{-G}\left(-{1 \over 2}G^{\mu\nu}
\partial_\mu\phi\partial_\nu\phi - {1 \over 2}m^2\phi^2\right)\ .
\ee
One assumes that there are branes at $t=0$ and $t=T$, whose 
action is given by
\be
\label{iii}
S_0=\int\sqrt{-g}V_0(\phi)\ , \quad
S_T=\int\sqrt{-g}V_T(\phi)\ .
\ee
where $-T\leq t \leq T$ and  $t$ is identified with $-t$.
Note that in the same way as in AdS bulk one can identify $T$ 
with the vacuum expectation value of temporal radion field. 
In fact, $T$ is a dynamical variable, which can be included in 
the $(tt)$-component $G_{tt}$ of the metric tensor in (\ref{i}). 
If we change $G_{tt}$ as $C^2G_{tt}$ by a positive constant $C$, 
the last expression in (\ref{i}) is changed as $ds^2 = 
 - C^2dt^2 + \e^{-2kC|t|}\sum_{i=1}^4\left(dx^i\right)^2$. 
Further rescaling  the time coordinate $t\to {t \over C}$, 
we obtain the last expression in (\ref{i}), again, but the region 
where the coordinate $t$ takes its value is changed as 
$-{T \over C}\leq t \leq {T \over C}$.  Repeating back this process, 
we find that 
$T$ is a dynamical variable. In the following, we solve the 
equation of motion for fixed $T$ and after that we substitute 
the obtained solution into the action (\ref{ii}), where
 all the dynamical variables except $T$ are integrated out. 

With the assumption that the bulk de Sitter space is a background, by the 
variation over the scalar field $\phi$, we obtain the following 
equation:
\be
\label{iv}
0=\partial_\mu\left(\sqrt{-G}G^{\mu\nu}\partial_\nu\phi\right) 
 - m^2 \sqrt{-G}\phi - \sqrt{-g}V_0'(\phi)\delta(t) 
 - \sqrt{-g}V_T'\delta(t-T)\ .
\ee
Considering that $\phi$ only depends on $t$ and using (\ref{i}), 
one can rewrite (\ref{iv}) as 
\be
\label{v}
0=-\partial_t\left(\e^{-4k|t|}\partial_t\phi\right) 
 - m^2 \e^{-4k|t|}\phi - \e^{-4k|t|}V_0'(\phi)\delta(t) 
 - \e^{-4k|t|}V_T'\delta(t-T)\ .
\ee
In the bulk ($0<t<T$), the solution is given by
\be
\label{vii}
\phi = \e^{2kt}\left(a\e^{\sqrt{4k^2 - m^2} t} 
+ b\e^{-\sqrt{4k^2 - m^2}t}\right)\ .
\ee
Here $a$ and $b$ are constants of the integration. 
The solution (\ref{vii}) is similar to that for the case 
that the bulk is AdS. Indeed, for the space-like radial 
coodinate as $r$, the solution in the AdS bulk is given by
\be
\label{viiAdS}
\phi_{\rm AdS} = \e^{2kt}\left(a\e^{\sqrt{4k^2 + m^2} r} 
+ b\e^{-\sqrt{4k^2 + m^2}r}\right)\ .
\ee
The main difference is the sign in front of $m^2$ in the 
root. Then in the dS bulk, the solution behaves as an 
exponential function if $4k^2>m^2$ but if 
$4k^2<m^2$ the solution behaves with 
vibration:
\be
\label{viiV}
\phi = \e^{2kt}\left(a\e^{i\sqrt{m^2-4k^2} t} 
+ b\e^{-i\sqrt{m^2-4k^2}t}\right)\ .
\ee
Here $b$ should be a complex conjugate of $a$.
If $4k^2=m^2$, the solution has the following form:
\be
\label{viiA}
\phi=\left(a_1 + a_2 t\right)\e^{2kt}\ .
\ee

The constants $a$ and $b$ in (\ref{vii}) can be
determined to satisfy the boundary condition at $t=0$, $T$, 
coming from the $\delta$-functions in (\ref{v}):
\be
\label{viii}
\left.\partial_t\phi\right|_{t\to +0} - \left.\partial_t
\phi\right|_{t\to -0}=-V_0'(\phi)\ ,\quad
\left.\partial_t\phi\right|_{t\to T+0} - \left.\partial_t
\phi\right|_{t\to T-0}=V_T'(\phi)\ ,
\ee
that is,
\bea
\label{ix}
2\left\{\left(2k+\sqrt{4k^2 - m^2}\right)a 
+\left(2k-\sqrt{4k^2 - m^2}\right)b\right\}&=&V_0'(\phi)\ ,\nn
2\left\{\left(2k+\sqrt{4k^2 - m^2}\right)a \e^{
\left(2k+\sqrt{4k^2 - m^2}\right)T} \right. && \nn
\left. +\left(2k-\sqrt{4k^2 - m^2}\right)b\e^{
\left(2k-\sqrt{4k^2 - m^2}\right)T}
\right\}&=&-V_T'(\phi)\ .
\eea
By substituting the solution (\ref{vii}) into the action 
(\ref{ii}), one has
\bea
\label{x}
&& S_\phi=V_4\left[{\left(2k+\sqrt{4k^2 - m^2}\right)^2 -m^2
\over 4\sqrt{4k^2 - m^2}}a^2\left(\e^{2\sqrt{4k^2 - m^2}T} -1
\right) \right.\nn
&& \ \ \left. - {\left(2k-\sqrt{4k^2 - m^2}\right)^2 - m^2
\over 4\sqrt{4k^2 - m^2}}b^2\left(\e^{-2\sqrt{4k^2 - m^2}T} -1
\right) -2m^2 ab T\right]\ .
\eea
Here $V_4$ is the volume of the brane. 
By combining $S_\phi$ (\ref{x}) with $S_0(\phi_{t=0})$ and 
$S_T(\phi_{t=T})$ in (\ref{iii}), we might obtain the 
effective action for $T$. 
As a special model, we consider the case that 
\be
\label{xi}
V_0'=V_T'=\alpha\ (\mbox{constant})\ .
\ee
Then one can solve (\ref{ix}) with respect to $a$ and $b$:
\bea
\label{xii}
a&=&{\alpha\left(2k - \sqrt{4k^2 - m^2}\right)\left(
\e^{\left(2k - \sqrt{4k^2 - m^2}\right)T} + 1\right)
\over 2m^2\e^{2kT}\left(\e^{- \sqrt{4k^2 - m^2}T}
 - \e^{ \sqrt{4k^2 - m^2}T}\right)} \nn
b&=&-{\alpha\left(2k + \sqrt{4k^2 - m^2}\right)\left(
\e^{\left(2k + \sqrt{4k^2 - m^2}\right)T} + 1\right)
\over 2m^2\e^{2kT}\left(\e^{- \sqrt{4k^2 - m^2}T}
 - \e^{ \sqrt{4k^2 - m^2}T}\right)} \ .
\eea
By substituting (\ref{xii}) into (\ref{x}), we obtain
\bea
\label{xiii} 
&& S_\phi={\alpha^2 \e^{-4kT}V_4 \over 16m^2 \left(
\e^{- \sqrt{4k^2 - m^2}T} - \e^{ \sqrt{4k^2 - m^2}T}\right)^2} \nn
&& \ \times \left[{\left(m^2 - \left(2k-\sqrt{4k^2 - m^2}
\right)^2 \right)\left(\e^{\left(2k - \sqrt{4k^2 - m^2}\right)T} 
+ 1\right)^2\over \sqrt{4k^2 - m^2}} \right.\nn
&& \ \times \left(\e^{2\sqrt{4k^2 - m^2}T} -1\right)
 - \left(m^2 - \left(2k+\sqrt{4k^2 - m^2}
\right)^2 \right) \nn
&& \times {\left(\e^{\left(2k + \sqrt{4k^2 - m^2}\right)T} 
+ 1\right)^2\left(\e^{-2\sqrt{4k^2 - m^2}T} -1\right)
\over \sqrt{4k^2 - m^2}} \nn
&& \ \left. +8m^2 T \left(\e^{\left(2k - \sqrt{4k^2 - m^2}\right)T} 
+ 1\right)\left(\e^{\left(2k + \sqrt{4k^2 - m^2}\right)T} 
+ 1\right)\right]\ .
\eea
One can regard $-S_\phi$ as an effective potential for radion $T$: 
\be
\label{xiv}
V(T)=-S_\phi\ .
\ee
Then when $T\rightarrow 0$, $V(T)$ behaves as 
\be
\label{xv}
V(T)\rightarrow {\alpha^2 V_4\left(2k^2 - m^2\right) \over 
m^2\left(4k^2 - m^2\right)}{1 \over T}\ .
\ee
On the other hand, when $T\rightarrow +\infty$, 
\be
\label{xvi}
V(T)\rightarrow {\alpha^2 V_4 \over 16m^2}\left[ 
{8k^2 - 2m^2 + 4k\sqrt{4k^2 - m^2} \over \sqrt{4k^2 - m^2}}
 - 8m^2 T \e^{- 2 \sqrt{4k^2 - m^2}T} \right]\ .
\ee
Note that $0< T < + \infty$. Eqs.(\ref{xv}) and (\ref{xvi}) 
tell 
that the potential is bounded below and has non-trivial minimum 
if $2k^2>m^2$. In order to generate the hierarchy between the Planck 
scale $10^{19}$ GeV and weak scale $10^2$ GeV, if we assume 
$kT\sim 50$ and also 
$\sqrt{4k^2 - m^2}T \sim 1$, we can approximate the potential 
by (\ref{xvi}), which has a minimum at 
\be
\label{xvii}
T\sim {1 \over 2\sqrt{4k^2 - m^2}}\ .
\ee
Then in order to generate the hierarchy the fine-tuning is
\be
\label{xviii}
4-{k^2 \over m^2}\sim 10^{-4}\ ,
\ee
which might be  natural if compare with the ratio of the 
weak and the Planck scale $10^2/10^{19}=10^{-17}$. 

Now as the brane is Euclidean, in order to relate the above 
theory with the real electroweak physics, we need to 
Wick-rotate one more spatial coordinate on the brane. Then 
we have two time coordinates, which might cause the problem 
with the unitarity. The two time theory might not, at present, 
be so unnatural as the so-called F-theory \cite{Vafa} is realization of  
 two time theory. Even in the model here, the de Sitter 
time might be a rather formal one. 

Another way to relate the above model with 
the real electroweak theory is that one regards a spatial 
coordinate, say $x^4$ in (\ref{i}) as an imaginary time 
with a period $\beta={1 \over T_m}$. 
Then the model under discussion 
 describes the theory at finite temperature $T_m$. 
Then the CFT at finite temperature may be naturally 
described in the present framework. Furthermore if we 
put $ds^2=0$ in (\ref{i}) by choosing $x^1=x^2=x^3=0$, we
obtain $dt=\pm \e^{-kt}dx^4$, that is, 
\be
\label{sgmt1}
x^4 - x^4_0 = \pm {\e^{kt} \over k}\ .
\ee
Here $x^4_0$ is a constant of the integration. Since the 
temperature of the universe is proportional to the inverse 
of the scale of the universe,  Eq.(\ref{sgmt1}) indicates that 
the temperature decreases exponentially with time.

On the other hand, one can consider the case that the time 
coordinate is unique. Then if identify the time $t$ with $-t$, 
the obtained solution  
describes that the brane decays at $t=0$ 
into two bulk spacetimes (the bulk spacetimes are pair-created) 
or the bulk spacetimes are pair-annihilated  creating a brane 
at $t=T$ as is shown in Fig.\ref{Fig1}.

\unitlength=0.5mm

\begin{figure}
\begin{picture}(100,100)
\thicklines
\put(50,10){\circle*{4}}
\put(50,90){\circle*{4}}
\qbezier[100](50,10)(30,20)(30,50)
\qbezier[100](50,10)(70,20)(70,50)
\qbezier[100](50,90)(30,80)(30,50)
\qbezier[100](50,90)(70,80)(70,50)
\put(30,49){\vector(0,1){2}}
\put(70,49){\vector(0,1){2}}
\put(40,3){$t=0$}
\put(40,92){$t=T$}
\end{picture}
\caption{The pair-creation and the pair-annhilation by the 
space-like branes at $t=0,T$. }
\label{Fig1}
\end{figure}
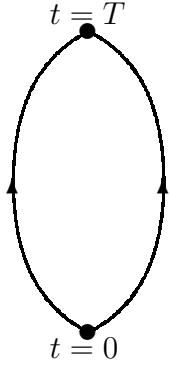

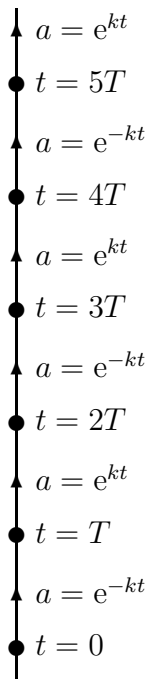
\begin{figure}
\begin{picture}(100,200)
\thicklines

\put(50,10){\circle*{4}}
\put(50,40){\circle*{4}}
\put(50,70){\circle*{4}}
\put(50,100){\circle*{4}}
\put(50,130){\circle*{4}}
\put(50,160){\circle*{4}}

\put(50,2){\line(0,1){8}}
\put(50,10){\vector(0,1){17}}
\put(50,27){\line(0,1){13}}
\put(50,40){\vector(0,1){17}}
\put(50,57){\line(0,1){13}}
\put(50,70){\vector(0,1){17}}
\put(50,87){\line(0,1){13}}
\put(50,100){\vector(0,1){17}}
\put(50,117){\line(0,1){13}}
\put(50,130){\vector(0,1){17}}
\put(50,147){\line(0,1){13}}
\put(50,160){\vector(0,1){17}}
\put(50,177){\line(0,1){3}}

\put(55,22){$a=\e^{-kt}$}
\put(55,52){$a=\e^{kt}$}
\put(55,82){$a=\e^{-kt}$}
\put(55,112){$a=\e^{kt}$}
\put(55,142){$a=\e^{-kt}$}
\put(55,172){$a=\e^{kt}$}

\put(55,8){$t=0$}
\put(55,38){$t=T$}
\put(55,68){$t=2T$}
\put(55,98){$t=3T$}
\put(55,128){$t=4T$}
\put(55,158){$t=5T$}

\end{picture}
\caption{The universe repeats shrinking and expansion 
with the period of $2T$ by the brane. }
\label{Fig2}
\end{figure}

If we do not do the identification, the universe has a 
periodicity of $2T$ as shown in Fig.\ref{Fig2}. When 
$0<t<T$, the scale of the universe shrinks as $a=\e^{-kt}$ 
and when $T<t<2T$, the universe expands and $a=\e^{kt}$. When 
$2T<t<3T$, as $a=\e^{-kt}$ and $3T<t<4T$, as $a=\e^{kt}$, etc.  
In this senario, in order to realize  such 5-dimensional 
universe, which repeats shrinking and expansion, we need 
a matter to compensate the jump of the first derivative 
of the metric. The matter is is given by the space-like brane. 
Then our analysis indicates that the brane can decay by the
effect of the scalar field but the brane could be created again, 
the created brane decays again. The period of repeating the
shrinking and expansion depends on the details of the scalar
part of the action. 

In the usual time-like branes in the AdS bulk spacetime, 
the branes can propagate in the time direction with a 
finite velocity, keeping the mutual distance $R$. 
The corresponding solution may exist in the dS bulk case with the
role of $t$ and $x^5$ exchanged.  Even if such a solution 
really exist, this would not change the above senarios 
drastically. 

Thus, our study indicates that if observable Universe is the brane 
embedded in five-dimensional de Sitter bulk then stabilization of fifth
(temporal) 
dimension may occur via the minimum of modulus effective potential. 
The analog of radion may be introduced and hierarchy problem may be 
solved as well. However, as radion is related now with the time 
the interpretation of the radion physics and its relation with electroweak
theory is not so clear as in AdS bulk. In particulary, even the evolution of 
the observable Universe may be drastically changed or second time 
coordinate should be introduced.
 Note finally that very recently (after the first version of 
this paper appeared), the localization of the graviton on 
the brane emmbedded in five-dimensional de Sitter bulk 
 was shown in \cite{56}. 


\ 

\noindent
{\bf Acknoweledgements} 

This research is supported in part by the 
Grant-in-Aid for Scientific Research on Prioirity Areas (B) 
of The Ministry of Education, Culture, Sports, Science and 
Technology (MEXT) (No. 13135101 and 13135208). 
One of authors (S.O.) acknoweledges support of the  Grant No. 13135101, 
 which made possible to finish this work during his visit to Japan.


\end{document}